\documentclass[ACS,STIX1COL]{WileyNJD-v2}
\articletype{ORIGINAL ARTICLE}

\received{xxx}
\revised{xxx}
\accepted{xxx}

\newcommand{\onlinecite}[1]{\hspace{-1 ex} \nocite{#1}\citenum{#1}} 

\usepackage{bm}

\begin{document}

\title{Shock Physics in Warm Dense Matter--a quantum hydrodynamics perspective
}

\author[1]{F. Graziani*}
\author[2,3]{Z. Moldabekov$^\dagger$}
\author[1]{B. Olson}
\author[4]{M. Bonitz}

\authormark{F. Graziani \textsc{et al}}

\address[1]{\orgdiv{}\orgname{Lawrence Livermore National Laboratory},  \orgaddress{94550 \state{Livermore, CA}, \country{USA}}}

\address[2]{\orgdiv{}\orgname{Center for Advanced Systems Understanding (CASUS)}, \orgaddress{D-02826 \state{G\"orlitz}, \country{Germany}}}
\address[3]{\orgdiv{} \orgname{Helmholtz-Zentrum Dresden-Rossendorf}, \orgaddress{ D-01328\state{ Dresden}, \country{Germany}}}

\address[4]{\orgdiv{Institut f\"ur Theoretische Physik und Astrophysik}, \orgname{Christian-Albrechts-Universit\"at zu Kiel}, \orgaddress{\state{Leibnizstra{\ss}e 15, 24098 Kiel}, \country{Germany}}}


\corres{*\email{graziani1@llnl.gov}\\$^\dagger$\email{z.moldabekov@hzdr.de}}

\abstract{Warm dense matter (WDM)--an exotic, highly compressed state of matter between solid and plasma phases is of high current interest, in particular for astrophysics and inertial confinement fusion. For the latter, in particular the propagation of compression shocks is crucial. 
The main unknown in the shock propagation in WDM is the behavior of the electrons since they are governed by correlations, quantum and spin effects that need to be accounted for simultaneously. Here we describe the shock dynamics of the  warm dense electron gas using a quantum hydrodynamic model. From the numerical hydrodynamic simulations we observe that the quantum Bohm pressure  induces shear force which weakens the formation and strength of the shock. In addition, the Bohm pressure induces an electron density response which takes the form of oscillations. This is confirmed by the theoretical analysis of the early stage of the shock formation. Our theoretical and numerical analysis  allows us to identify characteristic dimensionless shock propagation parameters at which the effect of the Bohm force is important.   
}


\maketitle


\section{Introduction}\label{sec:1}

Warm dense matter (WDM)--an exotic state on the border of plasma physics and condensed matter physics, e.g.~Refs.~\onlinecite{graziani-book,Fortov2016, moldabekov_pre_18, dornheim_physrep_18}, is currently a very active research field. Among the occurences are the interior of giant planets, e.g. \cite{Militzer_PRE_2021,schlanges_cpp_95,bezkrovny_pre_4, vorberger_hydrogen-helium_2007, militzer_massive_2008, redmer_icarus_11,nettelmann_saturn_2013}, brown and white dwarf stars \cite{saumon_the_role_1992, chabrier_quantum_1993,chabrier_cooling_2000}, and the outer crust of neutron stars \cite{Haensel,daligault_electronion_2009}. 
In the laboratory, WDM is being produced via laser or ion beam compression, or with Z-pinches, see Ref.~\onlinecite{falk_2018} for a recent review. 
Aside from dense plasmas, also many condensed matter systems exhibit WDM behaviour -- if they are subject to strong excitation, e.g. by lasers or free electron lasers~\cite{Ernstorfer1033,PhysRevX.6.021003}.
Among the important technological applications of WDM is indirect and direct drive inertial confinement pulsed power driven fusion  \cite{moses_national_2009,matzen_pulsed-power-driven_2005,hurricane_inertially_2016} where electronic quantum effects are important during the initial phase of the target implosion. Laser or x-ray driven ablated plasmas accelerate the shell and produce a set of shock waves that compress and heat the target.    

In addition to the role shocks play in laser, x-ray and pulsed power driven fusion targets, shocks play a crucial role in determining the equation of state of materials  at high pressures and densities. For lower pressures (less than a Megabar), static laboratory techniques such as diamond anvil cells are routinely used to obtain equation of state (EOS) data. Above a Megabar, however, high pressures need to be  generated in the target via transient shock or ramp loading. For high pressure EOS experiments, the facilities that are mostly used are the National Ignition Facility (NIF), Omega laser at the Laboratory for Laser Energetic and the Z-Machine at Sandia National Laboratories.  A well established technique for obtaining equation of state measurements for materials in the solid to warm dense matter regime is based on producing a shock through either a planar or spherical target. Using the Rankine-Hugoniot relations, only the driver and shock velocities are required to obtain a mechanical equation of state. \cite{Zeldovich_Raizer,benedict2014,hu_militzer_PhysRevLett.104.235003,kritcher_Nature.584.51}

Characteristic of all these diverse warm dense matter systems is the important role of electronic quantum effects, moderate to strong Coulomb correlations and finite temperature effects. Quantum effects of electrons are of relevance at low temperature and/or if matter is very highly compressed, such that the temperature is of the order of (or lower than) the Fermi temperature, for a recent overview, see Ref.~\cite{bonitz_pop_20}. 
Due to the simultaneous relevance of these effects, a theoretical description of WDM is difficult. Among the actively used approaches are quantum kinetic theory
\cite{balzer_pra_10,balzer_pra_10_2,schluenzen_20_prl,schluenzen_cpp_18}, quantum hydrodynamics \cite{zhandos_pop18, zhandos_cpp17_1d, bonitz_pre_13,zhandos_cpp_19}, and density functional theory simulations, e.g.~\cite{collins_pre_95,plagemann_njp_2012,witte_prl_17,zhang_pre_21}. The most accurate results for warm dense matter, so far, were obtained via first principle computer simulations such as quantum Monte Carlo (QMC) \cite{sign_cite,militzer_path_2000,filinov_ppcf_01,filinov_pss_00,schoof_cpp11,dornheim_physrep_18,filinov_pre15,schoof_prl15,dornheim_njp15,pierleoni_cpp19}.

However, QMC simulations, so far, only describe thermodynamic equilibrium situations, including static and dynamic properties \cite{dornheim_prl_18, hamann_prb_20}. On the other hand, first principles time-dependent approaches, such as quantum kinetic methods and time-dependent DFT (TD-DFT) only capture relatively short periods of time, on the femtosecond to picosecond scale.
At the same time, phenomena such as shock propagation and various instabilities (such as Rayleigh-Taylor instability) take place at significantly larger length and time scales that are currently inaccessible to the aforementioned \textit{ab initio} methods. Naturally, due to the large number of involved particles, hydrodynamic modelling is the standard approach for this type of problems and is well established in the realm of classical physics. However, a strong density gradient at the front of the shock, in combination with the quantum degeneracy of the plasma, can require taking into account  quantum non-locality which is neglected in classical hydrodynamics. Therefore, quantum generalizations of  hydrodynamics (QHD) are required. 

QHD equations can be derived using different approaches.  One can derive the many-body QHD equations using, as a starting point, the quantum kinetic equation governing the dynamics of the one-particle distribution in phase space (Wigner function). Similar to the classical case, the QHD equations follow from taking the moments of the distribution function in the kinetic equations \cite{Manfredi_2001, Manfred_2008, manfredi_fields_05}. For correlated systems, this approach requires to use advanced quantum collision integrals that go beyond perturbation theory (second Born approximation, quantum Landau equation),  \cite{bonitz_book,kremp_99_pre,kosse-etal.97}. In particular, non-ideality and dynamical screening effects should be included which is becoming feasible within the recently derived G1--G2 scheme \cite{schluenzen_20_prl,joost_prb_20}.
An alternative method is formulated in terms of the orbitals from Kohn-Sham density functional theory (KS-DFT) \cite{moldabekov_qhd_21,bonitz_pop_19, manfredi_fields_05}. In the KS-DFT, electronic non-ideality is included via the Kohn-Sham exchange-correlation (xc) potential, and the accuracy of the approach is solely determined by the quality of this potential\cite{KS65, doi:10.1146/annurev-physchem-040214-121420}. Although the KS orbitals are auxiliary quantities, the total density is physical and, in principle, can be calculated with high accuracy using the existing xc functionals which include finite temperature effects \cite{groth_prl17}. 
A fourth approach to formulating QHD is based on the orbital-free density functional theory (OF-DFT) \cite{zhandos_pop18}. The quality of the OF-DFT-based QHD simulations depends on the approximation used for the non-interacting free energy density functional. The strength of this formalism is that it is accurate in the weak perturbation regime due to connection of the force field to the density-density response function of electrons and the local field correction \cite{zhandos_pop18}. The latter is accurately known from recent Quantum Monte-Carlo simulations~\cite{dornheim_jcp_19-nn, PhysRevB.103.165102}. Additionally, the ongoing development of non-interacting energy density functionals could make OF-DFT-based QHD reliable beyond the linear response regime \cite{PhysRevX.11.011049}. Finally, Michta \cite{Michta2019} has recently taken the original approach of Madelung and extended the ansatz \cite{madelung_27} to many-body fermions by assuming a many-body wave function to consist of a product of anti-symmetrized, uncorrelated electrons in the form of plane-waves, with the density gradient used as an expansion parameter. He was able to obtain a many-body form of the QHD equation that included the Bohm potential for the average density, in addition to Thomas-Fermi and Dirac exchange pressure terms and Hartree effects.
For an up-to-date overview of QHD, see the review articles by Manfredi \cite{Manfredi2021} and Bonitz \textit{et al.} \cite{bonitz_pop_19, bonitz_pop_20}. 

Regardless of the theoretical basis, one common aspect of the various variants of QHD is the Bohm potential which provides an intuitive approach to quantum non-locality effects. 
Essentially, the Bohm potential represents the machinery behind the quantum tunnelling of electrons to a  classically inaccessible region \cite{BOHM1987321}. While the Bohm potential for the single electron case is known from the original works of Bohm \cite{bohm_pr_52_i, BOHM1987321}, von Weizs\"acker, and Madelung \cite{madelung_27}, the validity range of the direct transfer, e.g. \cite{manfredi_prb_01,manfredi_fields_05} of this result to the many-electron case, where the electron probability density is  replaced by the mean density, remained open. Only recently a  
qualitative and quantitative
understanding of the Bohm potential for many electrons was achieved, based on  \textit{ab initio} calculations of the microscopic many-electron Bohm potential \cite{moldabekov_qhd_21}. An important conclusion of Ref.~\cite{moldabekov_qhd_21} is that the force on the electrons generated by the Bohm potential can be of the same order of magnitude as the force generated by the Thomas-Fermi pressure. In the case of a strong density gradient the Bohm force can even exceed the latter. While Ref.~\cite{moldabekov_qhd_21} studied the model case of a harmonic density perturbation, here we concentrate on a specific experimentally relevant situation that is important in high-energy-density plasmas -- shock waves. 


In this work, we present the first results of quantum hydrodynamic modelling with an emphasis on the effect of the Bohm potential on the formation and propagation of a shock wave in WDM and quantum plasmas. The goal of this article is to better understand the role the Bohm potential plays in modifying the spatial structure of the shock. For this reason, even though we show the full QHD equations with many-body Coulomb forces and physical viscosity, the latter effects will be ignored in this paper so we can focus on the effect of the Fermi and exchange pressure terms in the spatial structure of the shock and how the Bohm force adds additional complexity. The Bohm force, which is the gradient of the Bohm pressure on the right hand side of the QHD momentum equation, involves first, second order, and third order  derivatives of density with respect to space. Therefore one may anticipate that the Bohm potential will play a key role, especially in the presence of a density discontinuity.
As we will see, this is indeed the case: the higher order density gradients, introduce an oscillatory force term into the momentum equation which may significantly alter the spatial structure of the shock and, in particular, soften the discontinuity.  

The article is organized as follows. Section 2 provides an overview of the QHD equations we will be using and presents a useful set of scaling relations that convert the QHD equations into dimensionless form that are dependent only on the Brueckner parameter. Section 3 summarizes the shock physics of classical fluids and outlines the approach we will take to describe the formation and propagation of shocks within QHD.  Section 4 will discuss the theoretical basis for weak shock formation and propagation quasi-analytically. Section 5 will discuss numerical results for a 1D shock tube problem consisting of high density (pressure) and low density (pressure) region separated by a smooth transition region. We conclude with a discussion in Sec.~\ref{s:discussion}.

\section{Quantum hydrodynamics for dense plasmas}\label{s:qhd}
We will treat the quantum plasma as an electron fluid immersed in a uniform neutralizing ion background. The quantum plasma will be assumed to be at temperatures where the Fermi energy $E_F$ is dominant over the classical thermal energy, i.e. $\Theta=k_BT/E_F \ll 1$.  We will assume the QHD equations to be of the Bloch form with an equation of continuity and momentum, with pressure terms coming from Fermi, exchange and Bohm contributions. Even though we write down, for reference, the full QHD equations including the relevant terms coming from physical viscosity and many-body electrostatic forces \cite{Murillo17}, we will drop them for the rest of the paper so we can focus on the spatial structure of shock formation and propagation in an electron fluid, with and without the Bohm term. 

For completeness, in one space dimension, the QHD equations with the pressure, Bohm, Hartree and viscosity terms included, can be written in in conservative form as follows. 
\begin{align}\label{0}
\frac {\partial n}{\partial t} + \left(\frac{\partial n v}{\partial x}\right) &= 0~,\\[1ex]
\label{eq:1}
  \frac {\partial n v}{\partial t} + \left( \frac{\partial n {v^2}  }{\partial x}\right) &= -{\frac{1}{m}}\frac{\partial }{\partial x}\left(P_e + P_Q + P_X\right) +{\frac{e n}{m}}\frac{\partial \phi}{\partial x} + \beta \frac{\partial^2 v} {\partial x^2}~,
\end{align}
The number density and velocity of the quantum plasma are represented by $n$ and $v$ respectively. In Eq.~(\ref{eq:1}), the pressure term ($P_e$) coming from Fermi degeneracy, the Bohm potential contribution \cite{zhandos_pop18} ($P_Q$) and the exchange term ($P_X$) have the following forms: 
\begin{align}\label{2}
    P_e &= {\hbar^2}\frac{\left(3\pi^2\right)^{2/3} {n^{5/3}}}{5 m},\\[2ex]
    P_X &= -\left(\frac{3}{\pi}\right)^{1/3}\frac{e^2 {n^{4/3}}}{4},\\[2ex]
    P_Q &= -\gamma\,\frac{ {\hbar^2}}{4 m} n \frac{\partial^2 \log\left(n\right)} {\partial x^2}.
    \label{3}
\end{align}
%

Several comments are in order regarding the form of the QHD equations we have written down. First, as noted previously, the equations we use are of the Bloch form that were subsequently generalized by Ying \cite{1974_Ying}.  Second, we include viscosity terms in a manner fashioned after Diaw and Murillo \cite{Murillo17}. Using dynamic density functional theory \cite{Marconi99, Lutsko08, Diaw15} Diaw and Murillo \cite{Murillo17} generalized the QHD equations to include viscous terms.  Third, there is no assumption of incompressibility in spite of the fact that the viscosity term is proportional to the Laplacian of the velocity alone and not the Laplacian plus the gradient of the divergence of the velocity. This is because our QHD equations are in 1D. In 2D or 3D, there are two terms contributing to the viscosity. One term involves a Laplacian and  the coefficient of this term is the shear viscosity. The second term is the gradient of the divergence. In general, unless the fluid is incompressible, this term is non-zero. The coefficient in front of this term involves a sum of the bulk and shear viscosities. In 1D, which we consider in the paper, the two terms become simply the second derivative of the velocity with respect to space. The coefficient in front of the term is a sum of both bulk and shear viscosities. Fourth, we have chosen a relatively simple form of the exchange term. We will see in our simulations that the Dirac exchange is relatively small contribution when compared to the TF pressure and Bohm terms and hence we do not feel it will have a large effect on our results.  
Finally, the form of the Bohm potential used in Eq.~(\ref{3}) is appropriate for the conserved form of the QHD equations and is equivalent to that used in Moldabekov et al. \cite{zhandos_pop18} in the case that $\gamma=1$. In Ref. \cite{zhandos_pop18}, different $\gamma$ values were derived for different limiting cases (frequency-wave number combinations) assuming that the system is near equilibrium.
In contrast, the formation and propagation of a shock is a non-equilibrium process that is characterized by a significant density inhomogeneity. Therefore, the analysis performed in Ref. \cite{zhandos_pop18} cannot be directly applied to shock physics. How to extend the analysis of that reference to nonequilibrium is presently an open question. 
The aim of the present work is to establish the \textit{general features} of shock propagation in quantum plasmas arising due to the introduction of higher-order spatial derivatives of the density via the Bohm potential.  Therefore, we use the standard expression defined by Eq.~(\ref{3}). The results will need to be revised when a more accurate potential will be available for the description of the shock formation and propagation. 

 The Poisson equation which represents the mean-field $\phi$ in Eq.~(\ref{eq:1}) or Hartree energy  associated with the many-body electron system reads:
\begin{equation}\label{1a}
 \frac{\partial^2 \phi} {\partial x^2} = - 4 \pi e \left(n-n_{\rm ion}\right),
\end{equation}
where $n_{\rm ion}$ is the ion number density which for the simulations will be taken to be a uniform background. 
%
 In Eq.~(\ref{3}), $\gamma$ is a constant that is in the range between $0.1\dots 1$, an  analysis can be found in Ref.~\cite{zhandos_pop18}. 
Here we use a fixed value, $\gamma=1$, which is not critical for the main conclusions of the impact of the Bohm term. A discussion of this choice is given in the Conclusion. 

For completeness, we have written down the full set of QHD equations even though we will be be ignoring for this study the Hartree and physical viscosity terms. These terms will be addressed in future work. The purpose of this work is to focus on the effects of the Bohm term on the shock front and isolate it. 

In order to simplify the numerical simulations, we derive a dimensionless form of the QHD equations by introducing dimensionless variables in time, length, number density, and velocity, using the plasma frequency $\omega_p$, Fermi velocity $v_F$, Thomas-Fermi screening length $\lambda_F$ and ion number density, $n_{ion}= n_0$, as scaling factors.
The Brueckner parameter which serves as the nonideality parameter, is defined as the ratio of the mean inter-particle distance, $d=\left(4/3 \pi n_0\right)^{-1/3}$, to the first Bohr radius $a_B$,
\begin{equation}\label{1alt1}
r_s = \frac{d}{a_B}=\left({\frac{3}{4\pi n_0}}\right)^{1/3} \frac{me^2}{\hbar^2}
\end{equation}
where
\begin{equation}\label{1alt2}
a_B = \frac{\hbar^2}{m e^2}\,,
\end{equation}
and  $m$ is the electron mass.

We now define the ion density as a function of $r_s$ and $a_B$
\begin{equation}\label{1alt3}
n_0 = \frac{1}{V_B r_s^3}\,,
\end{equation}
where $V_B=\frac{4\pi}{3} a_B^3$ is the Bohr volume.
We can now define velocity, length and time scales based on the Fermi velocity, Fermi length and plasma frequency:
\begin{equation}\label{1alt5}
v_{F_0}^2 = \alpha^2 c^2 \left(\frac{9\pi}{4}\right)^{2/3} \frac{1}{r_s^2},\qquad 
\lambda_{F_0}^2 = \frac{1}{4} \left(\frac{4\pi^2}{9}\right)^{1/3} a_B^2 r_s, \qquad 
{\omega_{p_0}^2 = 3 \alpha^2 \frac{c^2}{a_B^2}}\frac{1}{r_s^3}\,,\\
\end{equation}
where $\alpha=1/137.036$ is the fine structure constant. A useful identity $v_{F_0}/(\lambda_{F_0} \omega_{p_0}) = \sqrt{3}$  can be easily derived which helps in deriving the dimensionless form of the QHD equations.
Using these scaling factors we introduce the dimensionless time, length, and velocity
\begin{align}
     \xi & = x/\sqrt{3} \lambda_{F_0}\,,
     \nonumber\\
     \tau &= t \omega_{p_0}\,,\\
     \label{eq:density_unit}
     \eta(\xi,\tau) &= n(x,t)V_B r_s^3\,,\\
     \epsilon(\xi,\tau) &= v(x,t)/v_{F_0}\,.
\end{align}

Using the above scaling relations, we obtain the following dimensionless expressions for the conserved form of the continuity and momentum equations which will prove useful when we present the numerical simulation of shocks physics in QHD,

\begin{align}\label{11}
\frac {\partial \eta}{\partial \tau} + \frac{\partial \eta \epsilon}{\partial \xi} &= 0,\\
\label{12}
  \frac {\partial \eta \epsilon}{\partial \tau} + \frac{\partial\left(  \eta {\epsilon^2} \right) }{\partial \xi} &= -\frac{\partial \left(\Pi_e + \Pi_Q + \Pi_X\right)}{\partial \xi} +\eta \frac{\partial \Phi}{\partial \xi} + \tilde{\beta} \frac{\partial^2 \epsilon} {\partial \xi^2}\,,
\end{align}
where the dimensionless Fermi, exchange, Bohm pressures and viscosity are given by

\begin{equation}\label{13}
    \Pi_e= \frac{ {\eta^{5/3}}}{5}, \quad 
    \Pi_X= - C r_s \eta^{4/3}, \quad 
    \Pi_Q= -A \gamma r_s  \eta \frac{\partial^2 \log\left(\eta\right)} {\partial \xi^2},\quad
    \tilde{\beta}=\beta \frac{V_B {r_s}^3}{{v_F}^2}
\end{equation}
with $C=0.0416$ and $A = 0.0553$ being dimensionless constants.

The scaled Poisson equation for the dimensionless electrostatic potential ${\Phi}$, defined by $\phi = \frac{B}{r_s^2} \frac{e}{a_B}\Phi$ is given by

\begin{equation}\label{16}
\frac {\partial^2 {\Phi}}{\partial {{\xi}}^2} = {\eta}-1\,,
\end{equation}
where $B=3.68$. 

\section{Hydrodynamics of shock formation in the absence of quantum diffraction effects}\label{s:classical}
Shocks appear in solids, fluids, gases -- collisional and collisionless -- as the result of being subjected to a pressure drive coming from a piston, laser or high explosive. Shocks induce a sudden transition in the  velocity, pressure and density state of the material. The relationship between these states, post and pre-shock, are given by the Rankine-Hugoniot relations or jump conditions. Mathematically, the Euler equations admit discontinuous solutions in density, velocity and pressure. The shock solution is a consequence of a singularity in the classical Euler equations that is characterized by a steepening and eventual overtaking of the flow, signaling shock formation. The classical Euler equations can also be cast as a Riemann problem with a set of characteristics describing  the trajectories of small amplitude perturbations. In the event that the fluid is undergoing compression, these characteristics eventually intersect at a point. Intersecting characteristics is a signature of a multi-valued flow variables, indicating that a shock has formed   \cite{courant-friedrichs,landau-lifschitz,Atzeni2004, Drake2006}. Recently, there has been interest in the spatial structure of shocks and their formation \cite{Pomeau2008,Eggers2017,Ryutov2019}. The question arises of how  a shock emerges from an initially smooth flow. For the purposes of our study, we adopt the approach of Ryutov \cite{Ryutov2019} for both Euler hydrodynamics and QHD.

In order to prepare for the discussion of the formation and propagation of 1D shocks within QHD, we take a step back and consider the non-conservative form of the QHD equations for an ideal Fermi gas without exchange or Bohm terms. This means, we  neglect quantum diffraction effects leading to the model of a ``semi-classical Fermi gas''. The range of validity of this model will become clear from Eq.~(\ref{eq:gamma-def}) below.
The equations are basically just Euler equations with a Thomas Fermi equation of state (instead of the classical ideal gas equation of state) and no energy equation. The question is, do these equations exhibit shocks and if so, how do they form?

The simplified QHD equations read: 
\begin{align}\label{classical1}
\frac {\partial n}{\partial t} + \left(\frac{\partial n v}{\partial x}\right) &= 0~,\\[1ex]
\label{1}
  \frac {\partial v}{\partial t} + v \left( \frac{\partial v  }{\partial x}\right) &= -{\frac{1}{mn}}\frac{\partial P_e}{\partial x},
\end{align}
where the number density and velocity of the Fermi gas are represented by $n$ and $v$, respectively. 
We consider a co-moving frame where the flow is to the left, in the negative x direction with speed $c_s$. Similar to Ryutov \cite{Ryutov2019}, we consider a uniform sonic flow plus small deviations that lead to an overtaking and formation of a shock. Because the deviations are small, the results shown here pertain to weak shocks. The velocity of the perturbed flow is defined by,
\begin{align}\label{classical2}
v = -c_s + \delta v 
\end{align}
\begin{align}\label{classical3}
c_s=\left(\frac{5}{3} \frac{P_t}{m n_t}\right)^{1/2},
\end{align}
where $P_t$ and $n_t$ are the uniform pressure and number density in the vicinity of the shock transition appropriate for a degenerate Fermi gas given by,
\begin{equation}\label{classical2a}
    P_t= {\hbar^2}\frac{\left(3\pi^2\right)^{2/3} {{n_t}^{5/3}}}{5 m}\,. \quad 
\end{equation}
Using the Euler equations (\ref{classical1}) and expanding about the uniform density and velocity, we obtain to second order in the density and velocity perturbations, the following,
\begin{align}\label{classical4}
\frac {\partial \delta n}{\partial t} - c_s \frac{\partial \delta n}{\partial x} + \delta v \frac{\partial \delta n}{\partial x} + n_t \left(1 + \frac{\delta n}{n_t}\right) \frac{\partial \delta v}{\partial x} &= 0\,,
\\
\label{classical5}
  \frac {\partial \delta v}{\partial t} - c_s \frac{\partial \delta v}{\partial x} + \delta v \frac{\partial \delta v}{\partial x} &= -\frac{1}{m n_t}\left(1 - \frac{\delta n}{n_t}\right) \frac{\partial \delta P}{\partial x},
\\
\label{classical6}
  \delta P &= m{c_s}^2 \delta n \left(1 + \frac{\delta n}{3 n_t}\right)\,.
\end{align}

If we ignore the quadratic terms in equations (\ref{classical4}) and (\ref{classical5}), they exhibit a steady state solution corresponding to a linear wave co-moving with the fluid. This approximation effectively describes the fluid flow prior to shock formation occurring since the linearized equations will not support shock solutions. The solution is

\begin{align}\label{classical7}
  \delta n &= \frac{n_t}{c_s} \delta v,
\\
%
\label{classical8}
  \delta P &= \frac{5}{3} \frac{P_t}{c_s}\delta v\,.
\end{align}
Note that equations (\ref{classical7}) and (\ref{classical8}), when combined, yield the linear contribution to equation (\ref{classical6}), which is the full variation of the pressure due to density perturbations assuming a Thomas-Fermi equation of state. It is possible to combine equations (\ref{classical4}-\ref{classical6}) to obtain a single equation for $\delta v$ to second order in the variations. The derivation requires careful attention to detail. In particular the order of terms one has to keep or discard. Following Ryutov, we write the equation for the velocity perturbation as,  
\begin{align}\label{classical9}
  \frac {\partial \delta v}{\partial t} - c_s \frac{\partial \delta v}{\partial x} + \delta v \frac{\partial \delta v}{\partial x} + \frac{\delta n}{m{n_t}^2} \frac{\partial \delta P}{\partial x} &= -\frac{1}{m n_t} \frac{\partial \delta P}{\partial x}\,.
\end{align}
Our goal is to replace the $\delta P$ equation in the above equation with terms involving $\delta v$. For the $\delta P$ term on the left hand side of Eq.~(\ref{classical9}), we only need to use the lowest order linear variation contributing to $\delta P$ since it is already a second order term. Using equations (\ref{classical6}) and (\ref{classical7}) to lowest order, we obtain 

\begin{align}\label{classical10}
  \delta P \approx m {c_s}^2 \delta n \approx m c_s n_t \delta v
\end{align}

The right hand side is not as simple to deal with since we want to retain both first and second order terms in $\delta v$. We begin by taking the spatial derivative of equation (\ref{classical6}) and obtain,

\begin{align}\label{classical11}
\frac {\partial \delta P}{\partial x} &= m {c_s}^2 \frac{\partial \delta n}{\partial x} + \frac{2}{3} \frac{m{c_s}^2}{n_t} \delta n \frac{\partial \delta n}{\partial x} 
\end{align}

\begin{figure}
\begin{center}
\includegraphics[width=0.485\textwidth]{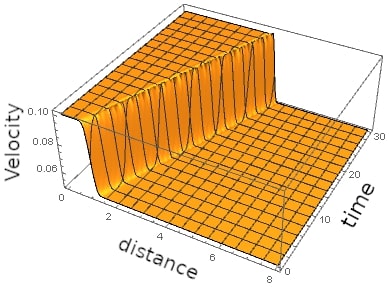}\hspace{0.015\textwidth}
\end{center}
\caption{Evolution and steepening of the velocity profile with the eventual formation of a shock. The units of velocity, distance and time are defined in Eq. (\ref{classical14}). }
\label{fig:name0}
\end{figure}

To ensure we have taken into all first and second order therms in $\delta v$, we use the continuity equation (\ref{classical4}) and re-express $\frac{\partial \delta n}{\partial x}$ as

\begin{align}\label{classical12}
\frac {\partial \delta n}{\partial x} &= \frac{1}{c_s} \left(\frac{\partial \delta n}{\partial t} + \delta v \frac{\partial \delta n}{\partial x} + n_t \frac{\partial \delta v}{\partial x} + \delta n \frac{\partial \delta v}{\partial x} \right)\,.
\end{align}
Substituting Eq.~(\ref{classical12}), into Eq.~(\ref{classical11}) and then into the right hand side of Eq.~(\ref{classical9}), we obtain

\begin{align}\label{classical13}
  \frac {\partial \delta v}{\partial t} + \frac{4}{3}\delta v \frac{\partial \delta v}{\partial x} &= 0\,.
\end{align}
This is a form of the inviscid Burgers equation and equivalent to Ryutov's equation \cite{Ryutov2019} with $\gamma=5/3$. It is well known that the solution to this equation can be constructed using the method of characteristics, i.e. straight lines determined by the initial data $v(x,0)$. Under certain initial conditions, this equation admits ``wave-breaking'' phenomena where the solution for $v$ starts to take on multiple values, signalling shock formation. Therefore, for a uniform sonic-flowing Fermi gas, initial velocity perturbations will evolve according to equation (\ref{classical13}) and shocks will form and propagate as it is illustrated in Fig.~ \ref{fig:name0}, where  we show the evolution of the velocity perturbations on the sonic flow with the initial conditions defined as 

\begin{equation}\label{classical14}
    \delta v= 0.05+0.025\left[1.0 - \tanh[10.0(x-1.0)]\right]
\end{equation}

The velocity perturbation described in Eq.~(\ref{classical14}), represents a weak deviation from a uniform flow velocity characteristic of what might be seen in a weak shock. 

\section{Shock formation in QHD: Theory}\label{s:qhdshocks1}
The presence of the Bohm pressure term has been of interest in the study of hydrodynamic instabilities in quantum plasmas \cite{Bychkov2008} where it was found that the the growth rate of the Rayleigh-Taylor instability could be reduced due to the presence of the Bohm term. Here we consider another application where the presence of the Bohm term could have significant impacts on the quantum plasma: a shock travelling in a quantum plasma exhibiting degeneracy, exchange effects, and quantum non-locality. The degeneracy and exchange pressures are associated with dense plasma effects, while the Bohm pressure is a measure of the quantum non-locality being sensitive to local density inhomogeneities and curvature.  The Bohm term is of particular interest because of its unusual functional dependence on the electron density. The Bohm pressure term is second order in the spatial derivative, and the corresponding Bohm force term is of third order. It is expected, therefore, that the Bohm contribution to the momentum equation will introduce oscillations whenever there is a discontinuity in the density profile of the quantum plasma. If large enough, the oscillations produced by the Bohm pressure could alter the shock propagation compared to the case when the Bohm term is not present. On the other hand, shocks travelling in a quantum plasma could give experimental evidence of the Bohm term.

Our starting point for developing a theoretical foundation of shock formation within QHD is the non-conservative form of the QHD equations with only the Bohm potential and Fermi electron pressure terms included for simplicity,

\begin{align}\label{quantum1}
\frac {\partial n}{\partial t} + \left(\frac{\partial n v}{\partial x}\right) &= 0~,\\[1ex]
\label{1}
  \frac {\partial v}{\partial t} + v \left( \frac{\partial v  }{\partial x}\right) &= -{\frac{1}{mn}}\frac{\partial P_e}{\partial x} - \frac{1}{m}\frac{\partial Q}{\partial x},
\end{align}
where we repeat the definition of the Fermi electron pressure (defined previously) and of  the Bohm potential as,

\begin{align}\label{quantum2}
    P_e &= {\hbar^2}\frac{\left(3\pi^2\right)^{2/3} {n^{5/3}}}{5 m},\\[2ex]
    Q &= -\gamma\,\frac{ {\hbar^2}}{2 m} \frac{1}{\sqrt{n}} \frac{\partial^2 \sqrt{n}} {\partial x^2}.
\end{align}

There is a hint that the Bohm term could be particularly interesting when considering density discontinuities of the sort that can give rise to shocks. To see this, consider an initial density profile of the form of a regularized density discontinuity,  

\begin{equation}\label{quantum3}
    n = 0.5+0.25[1.0 - \tanh[2.0(x-3.0)]\}.
\end{equation}

The density described by Eq.~(\ref{quantum3}) is an example of a regularized profile in a shock tube where the high density, hence pressure, region transitions to a low density, hence low pressure region. The steepness of the profile can be adjusted through the argument in the $\tanh$ function.

\begin{figure}
\begin{center}
\includegraphics[width=0.485\textwidth]{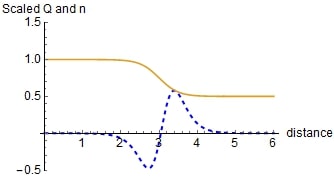}\hspace{0.015\textwidth}
\end{center}
\caption{The scaled Bohm potential  (Q divided by $\gamma \hbar^2/2m$, blue dashed line) for the density profile (in the units of $n_0$ defined in Eq. (\ref{1alt3}), full yellow line) shown in equation (\ref{quantum3}), normalized so the density is one at the origin. The dimensionless density is defined in Eq.~(\ref{eq:density_unit}).  Note that Q spans both negative and positive values, acting like an initial shearing force on the discontinuity, smearing it out. Both Q and n are normalized to highlight the functional dependence of Q on n. The distance is in the units of Thomas-Fermi screening length $\lambda_{F_0}$ defined in Eq. (\ref{1alt5}).} 
\label{fig:name000}
\end{figure}

The Bohm potential (up to constant factors), has the form shown in  Fig.~\ref{fig:name000}. It would be expected that, at least initially, this type of forcing function acting on the evolution of the density discontinuity would act as a shearing force centered in an area where the jump discontinuity is the largest, thus countering the steepening of the discontinuity and perhaps weakening or destroying the shock. Further, along with the numerical studies in the next section, we focus on the effect the Bohm term has on density discontinuities within QHD. 

In order to understand the formation of shocks within the QHD formalism in more detail, we again use the Ryutov approach \cite{Ryutov2019} and consider a uniform flow travelling with sonic velocity $c_s$ plus small deviations that lead to an overtaking and formation of a shock. Because the deviations are small, the results shown here pertain to weak shocks.We again define the velocity of the perturbed flow as,
\begin{align}\label{classical3a}
v &= -c_s + \delta v \,,
\\
c_s &=\left(\frac{5}{3} \frac{P_t}{m n_t}\right)^{1/2},
\end{align}
where $P_t$ and $n_t$ are the uniform pressure and number density in the vicinity of the shock transition appropriate for a degenerate Fermi gas. Expanding to second order in density and velocity perturbations, we obtain the the perturbed density and perturbed velocity equations,
\begin{align}\label{quantum4}
\frac {\partial \delta n}{\partial t} - c_s \frac{\partial \delta n}{\partial x} + \delta v \frac{\partial \delta n}{\partial x} + n_t \left(1 + \frac{\delta n}{n_t}\right) \frac{\partial \delta v}{\partial x} &= 0\,,\\
\label{quantum5}
  \frac {\partial \delta v}{\partial t} - c_s \frac{\partial \delta v}{\partial x} + \delta v \frac{\partial \delta v}{\partial x} &= -\frac{1}{m n_t}\left(1 - \frac{\delta n}{n_t}\right) \frac{\partial \delta P}{\partial x} -\frac{1}{m}\frac{\partial \delta Q}{\partial x}\,,\\
\label{quantum6}
  \delta P &= m{c_s}^2 \delta n \left(1 + \frac{\delta n}{3 n_t}\right),\\
  \delta Q = -\gamma\,\frac{ {\hbar^2}}{4 m n_t} \frac{\partial^2 \delta n} {\partial x^2} + \gamma\ \frac{ {\hbar^2}}{4 m {n_t}^2} \delta n \frac{\partial^2 \delta n}{\partial x^2} &+ \gamma\ \frac{ {\hbar^2}}{4 m {n_t}^2} \left(\frac{\partial \delta n}{\partial x}\right)^2,
  \label{quantum7}
\end{align}

Before we consider the full second order equations, we consider the linearized equations, which will hopefully provide insight into the properties of linear waves within QHD. Dropping terms quadratic in $\delta n$ and $\delta v$ , we obtain,

\begin{align}\label{quantum8}
\frac {\partial \delta n}{\partial t} - c_s \frac{\partial \delta n}{\partial x} + n_t \frac{\partial \delta v}{\partial x} &= 0
\end{align}

\begin{align}\label{quantum9}
  \frac {\partial \delta v}{\partial t} - c_s \frac{\partial \delta v}{\partial x} &= -\frac{{c_s}^2}{n_t} \frac{\partial \delta n}{\partial x} + \gamma\ \frac{ {\hbar^2}}{4 m n_t} \frac{\partial^3 \delta n}{\partial x^3}
\end{align}

Unlike the case of the QHD equations without the Bohm term, there is no obvious steady-state solution. We can gain additional insight into the nature of linear waves within QHD by looking for solutions of the form 

\begin{align}\label{quantum10}
\delta n = e^{i\omega t + i k x}A, \quad
\delta v = e^{i\omega t + i k x}B
\end{align}

Substituting equation (\ref{quantum10}) into equations (\ref{quantum8}) and (\ref{quantum9}), we obtain a scaling relationship between the density and velocity perturbations and the dispersion relation
\begin{align}\label{quantum11}
\delta n &= \frac{1}{1-\frac{\omega}{k c_s}} \, \frac{n_t}{c_s}\, \delta v\,,
\\
%
\label{quantum13}
\omega^2 &- 2 \omega k c_s -  \frac{\gamma {\hbar}^2 k^4}{4 m^2}= 0\,,
\end{align}
with the solution,
\begin{align}\label{quantum13}
\frac{\omega_{\pm}}{k c_s} = 1 \pm \sqrt{1 +\frac{\gamma {\hbar}^2 k^2}{4 m^2 {c_s}^2}}=1 \pm \sqrt{1 +\Gamma_k},
\end{align}
where we introduced the single-particle quantum diffraction parameter
\begin{align}
    \Gamma_k = \frac{\gamma {\hbar}^2 k^2}{4 m^2 {c_s}^2} = \frac{\gamma}{4}\frac{\epsilon(k)}{m c_s^2/2}\,,
    \label{eq:gamma-def}
\end{align}
which is defined as the ratio of the quantum kinetic energy of a particle, $\epsilon(k)=\frac{\hbar^2k^2}{2m}$, to the particle's kinetic energy of collective motion with the flow. From the solution (\ref{quantum13}) we learn that the parameter $\Gamma_k$ characterizes the relevance of quantum diffraction effects for shock physics.

\begin{figure}
\begin{center}
\includegraphics[width=0.485\textwidth]{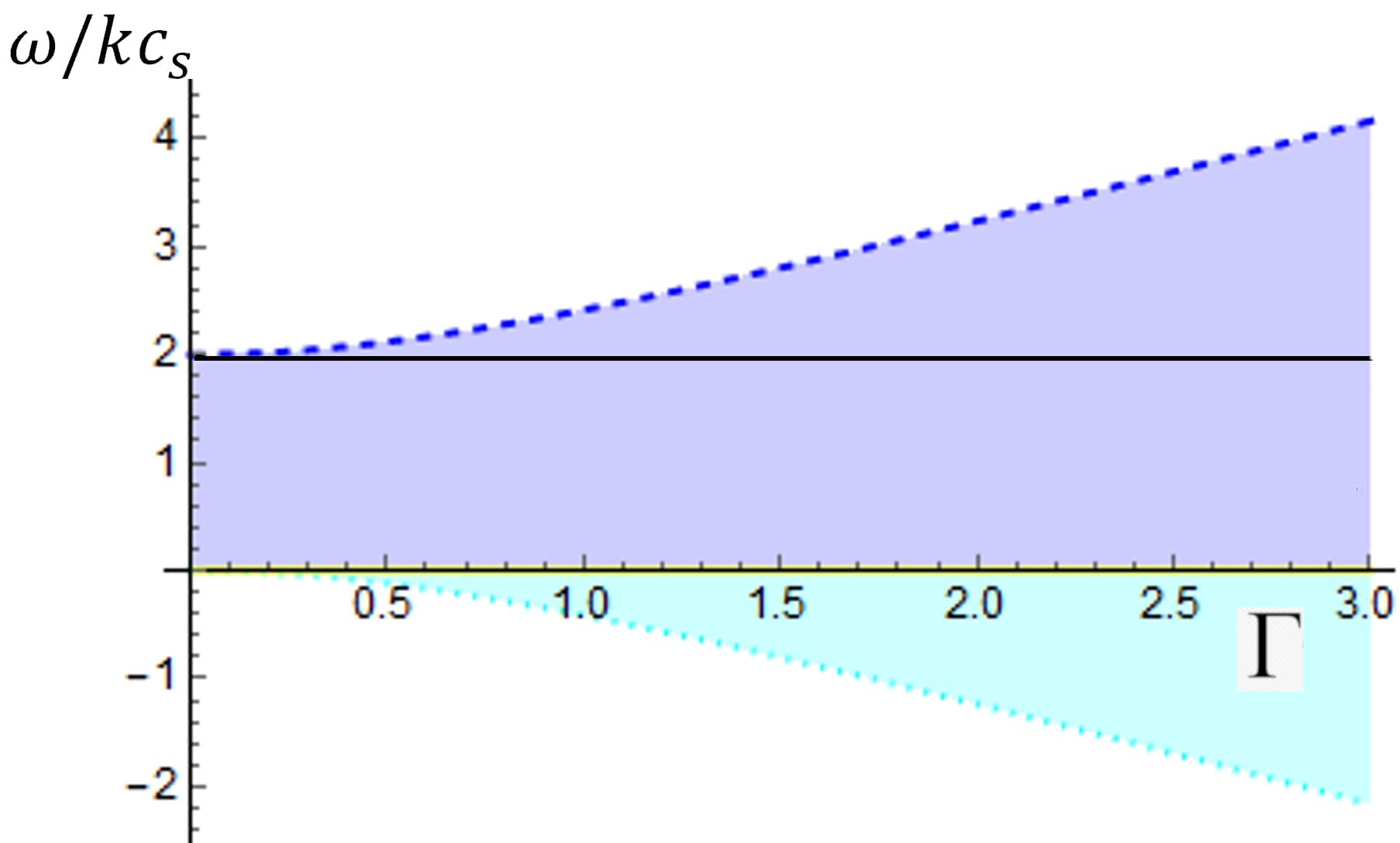}\hspace{0.015\textwidth}
\end{center}
\caption{Dispersion relation (\ref{quantum13}) as the function of the quantum diffraction parameter $\Gamma_k$ defined by Eq.~(\ref{eq:gamma-def}). The lines $\omega/kc_s=0$ and $\omega/kc_s=2$ (solid line) correspond to the perturbed velocity equations from semiclassical Fermi gas hydrodynamics without Bohm potential, i.e., $\gamma=0$. The  upper branch (dashed line) and lower negative branch (dotted line) correspond to the $\omega_{+}$ and $\omega_{-}$ solutions of Eq. (\ref{quantum13}), respectively.  
}
\label{fig:name00}
\end{figure}

We observe from Eq. (\ref{quantum13}) that, when $\gamma=0$, the dispersion relation has two solutions, $\omega=0$ and $\omega=k c_s$. This describes the semi-classical Fermi gas hydrodynamics we discussed in the previous section. The $\omega=0$ solution corresponds to the steady state while the solution $\omega= 2 k c_s$ describes a sound wave travelling with velocity $2 c_s$ in the reference frame of the shock. In this case we obtain the previously derived scaling relation solution $\delta n =\frac{n_t}{c_s} \delta v$. For $\gamma \not = 0$, we show in  Fig.~ \ref{fig:name00} a plot of both branches of the dispersion relation for the velocity perturbations in QHD.  We see that the steady state solution, characterized by $\omega=0$ is no longer a viable solution except for small wave numbers. However, we will still use the scaling relationship $\delta n =\frac{n_t}{c_s} \delta v$, assuming it is good enough, as long as $\Gamma_k \ll1$. 

Noticing that $\Gamma_k=(\gamma/4) k^2\Lambda_F^2$ and by making use of the fact that $c_s=v_F/\sqrt{3}$, the magnitude of the $\Gamma_k$ under the square root in the right hand side of Eq. (\ref{quantum13}) can be estimated using a type of deBroglie wavelength for an electron moving with the thermal velocity:

\begin{align}\label{quantum14}
\Lambda_F = \frac{\hbar}{m c_s}=a_B r_s\,.
\end{align}
As we will see in a moment, this term will set the spatial scale for the presence of the Bohm term.

Now that we have a sense of the properties of linear waves in QHD, we consider the perturbation equations (\ref{quantum4} - \ref{quantum7}), with quadratic terms included. We first use the continuity equation for the density perturbations to replace $\partial \delta n/ \partial x$ in the Fermi pressure and Bohm potential terms and we use the scaling relation from the continuity equation, $\delta n = n_t \delta v/c_s$

\begin{align}\label{quantum15}
\frac {\partial \delta n}{\partial x} = \frac{1}{c_s}\left[ \frac{\partial \delta n}{\partial t} + \delta v \frac{\partial \delta n}{\partial x} + n_t \left(1 + \frac{\delta n}{n_t}\right) \frac{\partial \delta v}{\partial x}\right] &= 0
\end{align}

The equation for the velocity perturbations now becomes

\begin{align}\label{quantum16}
  \frac {\partial \delta v}{\partial t} + \frac{4}{3} \delta v \frac{\partial \delta v}{\partial x}  &= \gamma\,\frac{ {\hbar^2}}{8 m^2 n_t c_s} \frac{\partial^2} {\partial x^2} \frac{\partial \delta v} {\partial t}+ \gamma\ \frac{ {\hbar^2}}{16 m^2 {c_s}^2}  \frac{\partial^2}{\partial x^2}\left(\delta v \frac{\partial \delta v} {\partial x}\right) + \gamma\ \frac{ {\hbar^2}}{8 m^2 {c_s}}\frac{\partial^3 \delta v} {\partial x^3} - \gamma\ \frac{ {\hbar^2}}{8 m^2 {c_s}^2}\left(\frac{\partial \delta v} {\partial x} \frac{\partial^2 \delta v} {\partial x^2}+\delta v \frac{\partial^3 \delta v} {\partial x^3}\right),
\end{align}

\begin{figure}
\begin{center}
\includegraphics[width=0.485\textwidth]{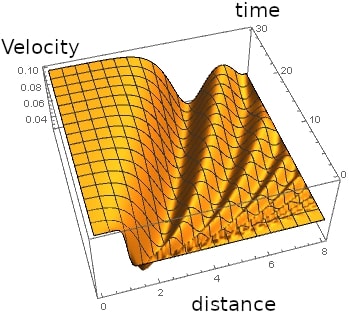}\hspace{0.015\textwidth}
\includegraphics[width=0.485\textwidth]{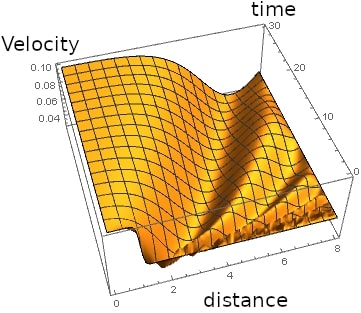}
\end{center}
\caption{Left: Evolution of the dimensionless velocity perturbations for  $r_s=0.5$ (left) and $r_s=1.0$ (right) as a function of scaled time (running from 0. to 30.) and scaled length (running from 0. to 8.).  Note the presence of oscillations coming from the Bohm terms. The wavelength of the oscillations increases with $r_s$, consistent with the length scale associated with the Bohm term. The units of velocity, distance and time are defined in Eq. (\ref{1alt5}).}
\label{fig:name01}
\end{figure}

We note that in the limit of $\gamma=0$ (i.e. no Bohm potential), we obtain the equation for the velocity perturbations in a semiclassical Fermi gas derived previously. Since all of the terms on the right hand side of equation (\ref{quantum16}) are proportional to $\hbar^2$ we can replace $\partial \delta v/\partial t$ on the r.h.s. of Eq.~(\ref{quantum16}) with equation (\ref{classical13}). The only error we are making involves  terms that are of higher order in $\hbar^2$. Equation (\ref{quantum16}) can be transformed further into a more transparent form if we define a dimensionless length z, time $\theta$ and velocity perturbation $\delta u$
\begin{align}\label{quantum17}
x &= \Lambda_F z, \quad
t = \frac{a_B \theta}{c_s}, \quad
\delta v = c_s \delta u\,,
\\
%
\label{quantum18}
  \frac {\partial \delta u}{\partial \theta} + \frac{4}{3} \delta u \frac{\partial \delta u}{\partial z}  &= \frac{7}{16}\gamma\ f_0 {r_s}^2  \frac{\partial \delta u} {\partial z} \frac{\partial^2 \delta u} {\partial z^2} + \frac{5}{48}\gamma\ f_0 {r_s}^2  \delta u  \frac{\partial^3 \delta u} {\partial z^3} + \frac{1}{8}\gamma\ f_0 {r_s}^2  \frac{\partial^3 \delta u} {\partial z^3}\,.
\end{align}
We note that the spatial scale for the Bohm term is determined by $\Lambda_F$ and the strength of the Bohm terms on the right hand side of Eq.~(\ref{quantum18}) is determined by the Brueckner parameter $r_s$. In Fig.~\ref{fig:name01} we see the numerical solution of equation (\ref{quantum18}) for $r_s=0.5$ and $r_s=1.0$. The initial and boundary conditions were identical to those used to solve the perturbed velocity equation in the previous section. We note the presence of oscillations in the perturbed velocity profile coming from the Bohm terms acting on the initial density profile and propagating as the discontinuity evolves. Note also that, unlike the steepening profile of the discontinuity and shock formation we saw for the semi-classical Fermi gas, no such phenomena occur here. The discontinuity is smeared out, preventing the formation of a shock. These observations will be confirmed in the next section where we solve the QHD equation numerically.

\section{Shock formation in QHD: Numerical results}
In order to better understand the physics of shock formation and propagation in quantum plasmas and to confirm the theoretical work discussed previously, we adapted the Miranda code and its Python interface Pyranda by adding QHD capabilities. Miranda is a multi-component, 1D-2D-3D radiation hydrodynamics code designed for large-eddy simulation of multi-component flows with turbulent mixing. It is capable of solving the multi-component compressible Navier-Stokes equations using adaptive mesh refinement. A variety of physics options are available, including radiation diffusion, magneto-hydrodynamics, self-gravity and thermonuclear fusion. The hydrodynamics package is based on tenth-order compact (Pade) schemes for spatial differencing, combined with fourth-order Runge-Kutta time-stepping.  Following standard practice, Miranda adds an artificial viscosity term to the momentum equation to control unphysical behavior due approximating the hydrodynamic equations on a mesh\cite{VonNeumann50, Bowers91}. The idea is to add purely numerical viscous terms that spread the shock front over several grid spacings thereby mitigating artificial ``ringing''  commonly seen in simulations of shock flows. The Miranda code has been used to model inter-facial instabilities (Rayleigh-Taylor, Richtmyer-Meshkov, Kelvin-Helmholtz), ICF implosions, ablation physics, supernovae, and classical fluids experiments (e.g., drop tanks, shock tubes, etc.) \cite{cook_2013,cook_2012,cook_2011,cook_2006}. The standard Miranda code was modified to allow for Fermi, exchange, and Bohm pressures. In the future, physical viscosity and a Hartree term will be added. 

We consider a simple initial 1D density profile spanning a density from $\eta=1.0$ to $\eta=2.0$ with a regularised discontinuity given by
\begin{equation}\label{16}
    \eta= 1.0 + 0.5\,\{1.0 + \tanh[W(\xi-\xi_0)]\}\,.
\end{equation}

This profile is used instead of a step function in order to better control numerical oscillations when the Bohm term is introduced. All simulations use the standard artificial viscosity (REF) to control numerical ringing at the steep density gradient. By varying W, we can simulate both ramp compression and shock compression experiments. A value of $W=2.0$ or $W=5.0$ produces enough of a sharp density drop that shocks form. 
The steepness W of the initial density profile was varied between 2.0 and 5.0. All simulations, unless mentioned otherwise, used a value of $W=5.0$. The initial position of the interface was also varied but for the purposes of this paper, it will be fixed and set at $\xi_0 = 7.0.$ The hydrodynamic mesh consisted of N zones spanning the 1D dimensionless length from $L=0.0$ to $L=12.0$. We ran simulations with $N=100, 200, 400, 800$ and $1000$ zones. The simulations shown here will be for $N=800$. The QHD equations were solved explicitly with a time step control restricted by the Courant condition. Simulations were typically run for over 50,000 cycles, until the initial profile had propagated and a shock had formed and propagated. The boundary conditions were fixed with the left most boundary density $\eta$ set to 1.0 and the right most boundary density set to 2.0. The dimensionless velocity is $\epsilon = 0.0$, at both the left and right boundaries.

\begin{figure}
\begin{center}
\includegraphics[width=0.445\textwidth]{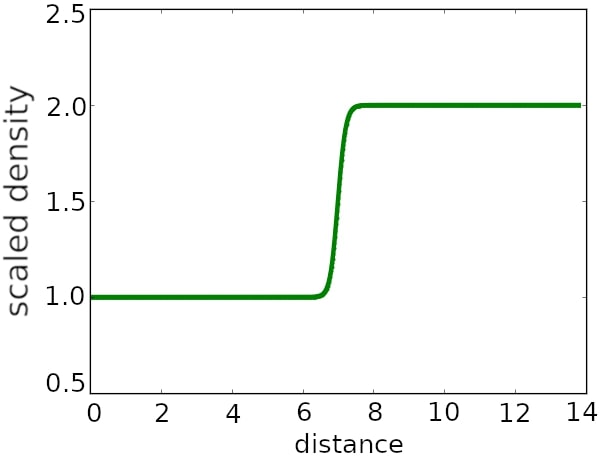}\hspace{0.015\textwidth}
\includegraphics[width=0.445\textwidth]{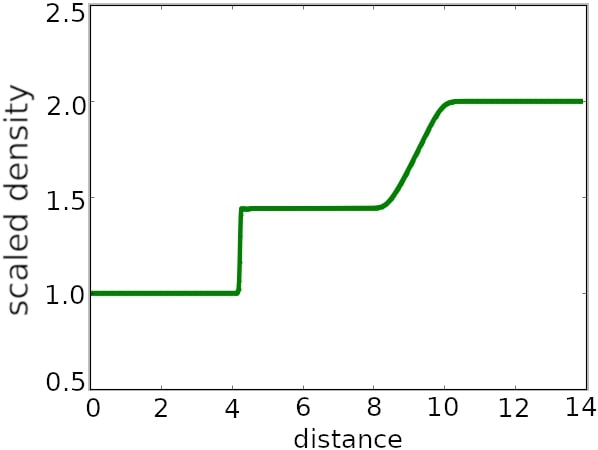}
\end{center}
\caption{Hydrodynamic simulations with Bohm and exchange pressure turned off (for details see text). Left: Initial density profile, Right: density profile at $t=3.77\, t_s$, where $t_s=(a_B/c_s)$. Note the shock formation and features reminiscent of shock physics for classical ideal fluids. Also notice the absence of any numerical ringing. The units of the distance and density are defined in Eq. (\ref{1alt5}) and Eq. (\ref{1alt3}), respectively.}
\label{fig:name1}
\end{figure}

In Fig.~\ref{fig:name1}, the initial and final density profiles for a standard hydrodynamic simulation of a shock are shown, albeit with Thomas-Fermi pressure. In order to produce these simulations, the Brueckner parameter was set to $r_s=0$ which effectively turns off Bohm and exchange pressure terms\footnote{This is not to be confused with the high-density limit.}. Unlike the classical simulation for shock phenomena which uses the classical ideal gas EOS, we emphasise that these shocks form using an ideal EOS for degenerate fermions. In all simulations shown an artificial viscosity is used to control ringing at sharp density variations \cite{VonNeumann50,Bowers91}. The simulation shown in figure \ref{fig:name1} will be our standard baseline for comparison and to assess the impact of the Bohm pressure term.

\begin{figure}
\begin{center}
\includegraphics[width=0.475\textwidth]{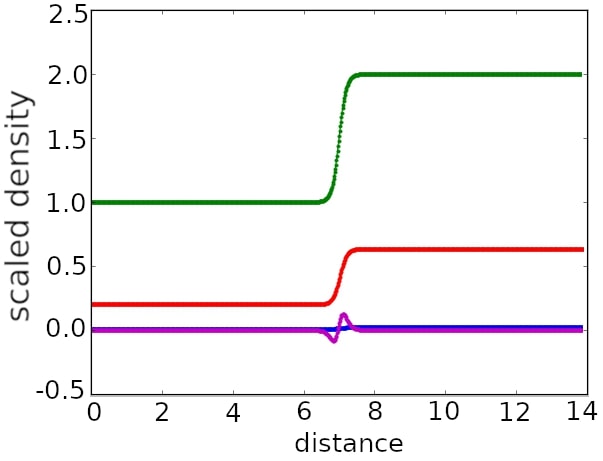}\hspace{0.015\textwidth}
\includegraphics[width=0.475\textwidth]{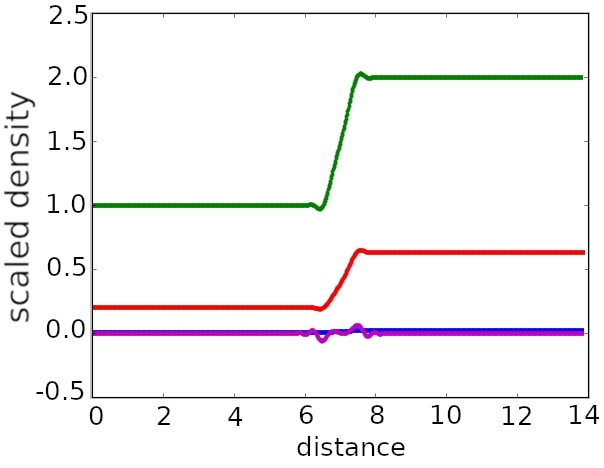}\\[1ex]
\includegraphics[width=0.475\textwidth]{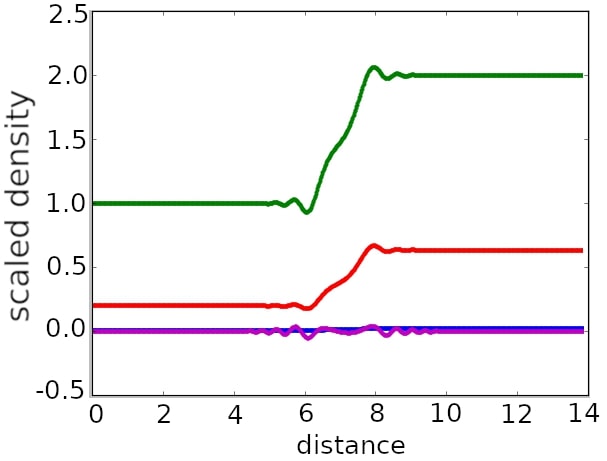}\hspace{0.015\textwidth}
\includegraphics[width=0.475\textwidth]{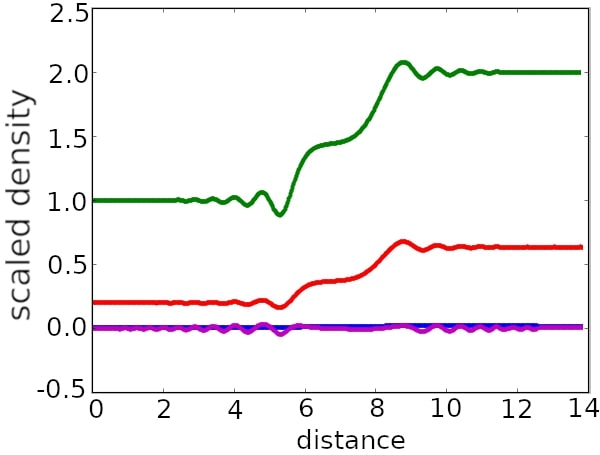}
\end{center}
\caption{Top left: Initial density profile (green). Top right: Density  profile at $t=0.25\,t_s$, where $t_s=(a_B/c_s)$. Bottom left: $t=0.62\,t_s$, bottom right: $t=1.51\,t_s$. The run corresponds to $r_s=0.5$
Red lines: Thomas-Fermi pressure, blue: exchange pressure,  and purple: the Bohm pressure. Note the oscillation present in the Bohm pressure at the initial moment. The units of the distance and density are defined in Eq. (\ref{1alt5}) and Eq. (\ref{1alt3}), respectively.
}
\label{fig:name2}
\end{figure}

\begin{figure}
\begin{center}
\includegraphics[width=0.475\textwidth]{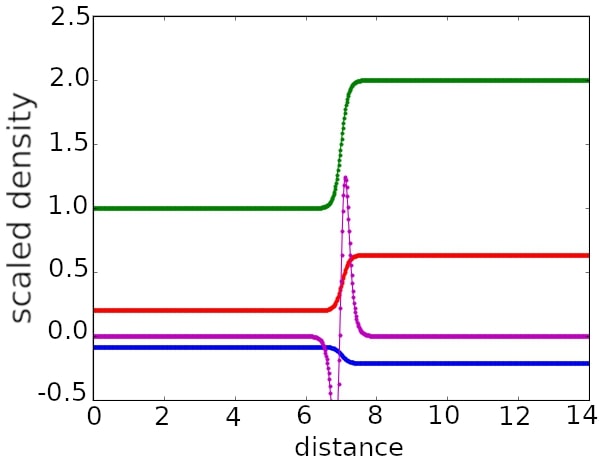}\hspace{0.015\textwidth}
\includegraphics[width=0.475\textwidth]{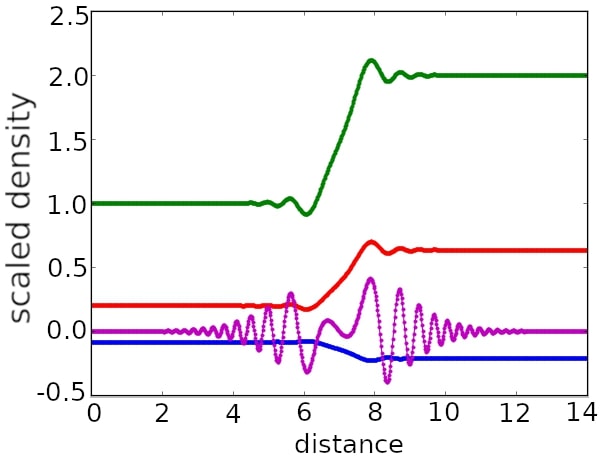}\\[1ex]
\includegraphics[width=0.475\textwidth]{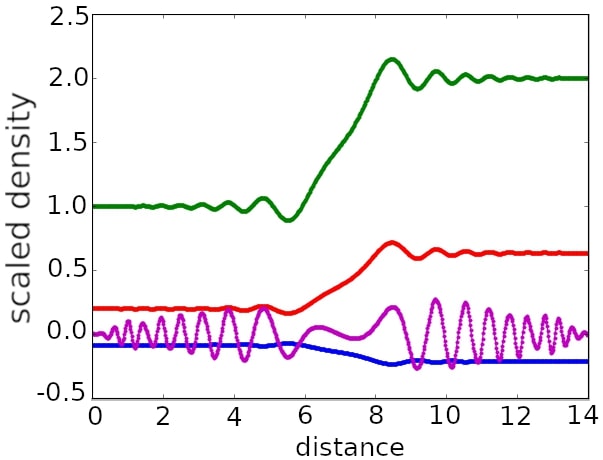}\hspace{0.015\textwidth}
\includegraphics[width=0.475\textwidth]{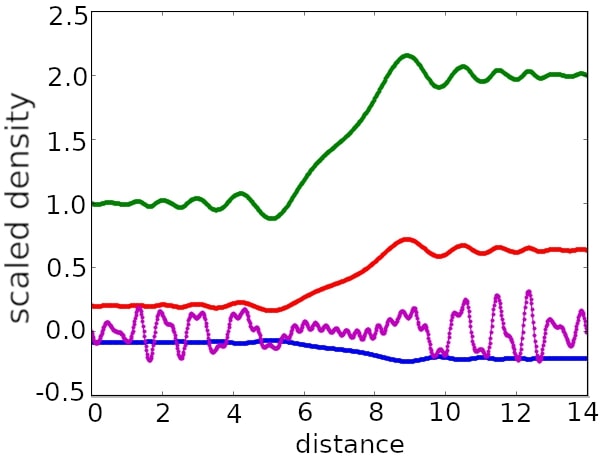}
\end{center}
\caption{Top left: Initial density profile (green). Top right: Density  profile at $t=0.25\,t_s$. Bottom left: $t=0.62\,t_s$, bottom right: $t=1.00\, t_s$. The unit of time is defined as $t_s=(a_B/c_s)$. The run is performed for $r_s=2.0$.
Red lines: Thomas-Fermi pressure, blue: exchange pressure,  and purple: the Bohm pressure. Note the oscillation present in the Bohm pressure at the initial moment. The units of the distance and density are defined in Eq. (\ref{1alt5}) and Eq. (\ref{1alt3}), respectively. 
}
\label{fig:name3}
\end{figure}


We now consider a full QHD simulation of a shock without the effects of the self-generated electric fields. We choose $r_s=0.5$ and $r_s=2.0$, spanning moderately to strongly coupled regimes. The "classical" case with with turned off Bohm potential was discussed above. Figure \ref{fig:name2} 
shows the simulation with the Bohm and exchange terms turned on. We see that initially, the Thomas-Fermi pressure dominates while the Bohm pressure is next in terms of scale, while the exchange pressure is small. Even though small in scale, the presence of the Bohm term has two effects. One, is the shear force that the gradient of the Bohm pressure generates which tends to spread the density profile out where it is steepest. This, in turn, tends to reduce the amplitude of the variations of the Bohm pressure. The second effect is due to the initial oscillations from the Bohm potential acting as a perturbation term to the momentum hydrodynamic equation. The hydrodynamic equations  propagate the disturbance to the density field by creating a non-monotonic density profile. In a feedback process, the modified density profile produces even more oscillations in the Bohm pressure, albeit with reduced amplitude. The overall effect is to weaken the formation and strength of the shock,  see Fig.~\ref{fig:name2}. As we see in Figure \ref{fig:name3}, these points are even more apparent in the simulations run with $r_s=2.0$.

\begin{figure}
\begin{center}
\includegraphics[width=0.485\textwidth]{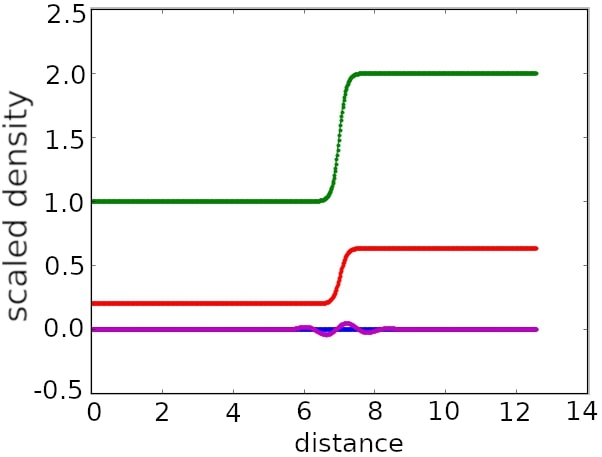}\hspace{0.015\textwidth}
\includegraphics[width=0.485\textwidth]{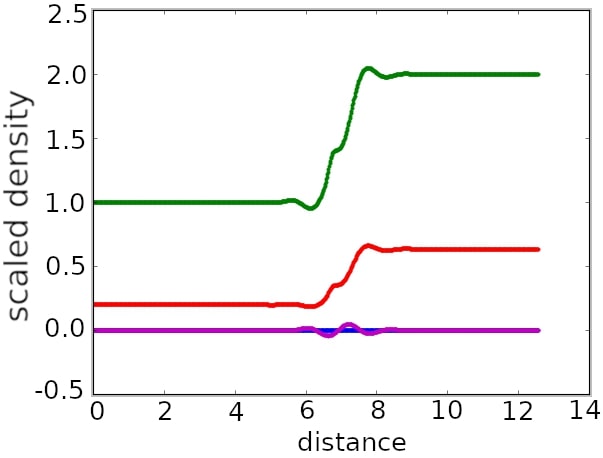}
\end{center}
\caption{Hydrodynamic simulation with the Bohm potential replaced by the static potential, Eq.~(\ref{eq:static-bohm}). Left: Initial state.
Right: $t=0.4\, t_s$. The unit of time is defined as $t_s=(a_B/c_s)$.  Green lines: density profile, red lines: Thomas Fermi pressure, blue lines: exchange pressure, and purple lines: static Bohm-type term (\ref{eq:static-bohm}). Note the presence of the non-monotonic density profile while a shock is starting to form. The units of the distance and density are defined in Eq. (\ref{1alt5}) and Eq. (\ref{1alt3}), respectively. $r_s$ is equal to 0.5.}
\label{fig:name4}
\end{figure}

We additionally test the observations that the oscillations we see in the simulations are physical and not generated numerically. For that we consider a test problem where the viscosity is set to zero and the Bohm pressure is replaced by an external perturbation defined by

\begin{equation}\label{eq:static-bohm}
    P_Q=0.05\cdot \sin[5.0 \cdot \xi]\cdot \exp[-(\xi - \xi_0)^2]
\end{equation}
This term mimics the initial oscillation in the Bohm pressure term in the presence of a jump in the density profile. It remains fixed in space for all time so it does not exactly replicate the impact of the Bohm pressure term but it does allow us to isolate the effect a local oscillating pressure term has on the evolution of the density profile. These test run is essentially just the baseline run discussed previously with the perturbation term $P_{Q}$ added to the Thomas-Fermi pressure. The results for this case  are presented in Fig.~\ref{fig:name4}. From  Fig.~\ref{fig:name4} we  observe that the effect of the fluctuating pressure perturbation on the propagation of  the density profile is similar to the impact of the Bohm pressure term presented in Figs.~\ref{fig:name2} and \ref{fig:name3}.
 Therefore, the Thomas-Fermi pressure term is driving the shock formation but the process is convolved with the formation of a non-monotonic density profile generated by the perturbation $P_{Q}$. Of course, since our perturbation is static, there is no feedback process between the changes in density and the changes in the external perturbation $P_{Q}$.
Nevertheless, this test reveals that that the oscillatory perturbation at the initial stage of the shock formation leads to the observed non-monotonic density profile when the Bohm term is taken into account. 


\section{Discussion and conclusions}\label{s:discussion}

In summary, we have studied the shock propagation in a warm dense electron gas at parameters relevant for WDM and ICF. 
The main focus of the presented study is the effect of the quantum non-locality, represented by the Bohm potential, on the  
formation and propagation of the shock front. The  analysis, based on the comparison of the results from  hydrodynamics simulations performed with and without Bohm potential, shows that the effect of the Bohm potential is in mitigating the shock front.
The mechanism behind this effect is the Bohm potential induced shear force which weakens the formation and strength of the shock.
Additionally, density oscillations appear in front and behind of the shock front. To confirm that these effects are not artifacts of numerical simulations, we have theoretically analysed the perturbative solution of the QHD equations with and without Bohm potential taken into account, up to the second order  in the velocity perturbation. This analysis has confirmed that the quantum non-locality results in a weaker shock and in the appearance of an oscillatory pattern around the shock front.   

In this work, the analysis was based on the standard Bohm potential derived by Bohm \cite{bohm_pr_52_i, BOHM1987321},  von Weizs\"acker, and Madelung \cite{madelung_27} for the single-electron case and does not take into account  many-particle effects.
It has recently been shown that this approximation is valid when the density perturbation is weak, $\delta n/n_0\ll 1$, and breaks down when the density perturbation is strong, $\delta n/n_0 \gtrsim 1$ \cite{moldabekov_qhd_21}, as is the case for shock propagation. However, it is not clear at the moment whether the modifications observed in Ref. will increase of decrease the trends reported in the present study. Nevertheless, the numerical study performed in the present paper is highly valuable since it provides a basic understanding of the effects produced by the presence of higher-order spatial derivatives   of the density, due to the Bohm potential in hydrodyanimcs. Moreover, from the presented numerical and analytical investigation, we identified characteristic shock parameters at which  electronic quantum non-locality are expected to become significant.  

Based on the findings presented in this work, further extension of the analysis of shock propagation in quantum plasmas is of high interest. This should explicitly take into account also the ionic dynamics and the Hartree mean field. Besides of that,  an improved Bohm potential for strongly perturbed  many-electron systems should be developed and parameterized as a function of density, based on the recent \textit{ab initio} simulations of Ref.~\cite{moldabekov_qhd_21}.  Implementing these results in the existing hydrodynamics codes should allow for an accurate and predictive modelling of shock propagation in quantum plasmas.       

\section*{Acknowledgements}
We acknowledge fruitful discussions with F. Beg, B. Cabot, A. Cook, T. Dornheim, D. Fratanduono, R. Kraus, R. Patterson, J. Vorberger.  Further, we acknowledge computing time at the Livermore Computing Center where simulations were performed. This work was performed under the auspices of the U.S. Department of Energy by Lawrence Livermore National Laboratory under Contract No. DE-AC52-07NA27344. MB has been supported by the Deutsche Forschungsgemeinschaft via grant BO1366/15-1. ZM acknowledges funding by the Center of Advanced Systems Understanding (CASUS) which is financed by the German Federal Ministry of Education and Research (BMBF) and by the Saxon Ministry for Science, Art, and Tourism (SMWK) with tax funds on the basis of the budget approved by the Saxon State Parliament. 

\bibliography{paper.bib}


\end{document}